\documentclass[twocolumn,aps,superscriptaddress,showpacs,preprintnumbers]{revtex4}

\pdfoutput=1
\usepackage[british]{babel}     
\usepackage{fancyhdr}    	
\usepackage[latin1]{inputenc}   

\usepackage{amsthm,amsfonts,amssymb,amsmath}
\usepackage{mathrsfs}
\usepackage{graphicx}
\usepackage{subfigure}
\usepackage{bm}			
\usepackage{longtable}

\usepackage{xcolor}

\usepackage{multirow}

\usepackage{graphicx}
\usepackage{epstopdf}

\usepackage{braket}		

\makeatletter
\renewcommand{\maketag@@@}[1]{\hbox{\m@th\normalsize\normalfont#1}}%
\makeatother

\def\r{\mathbf{r}}

\def\k{\mathbf{k}}
\def\q{\mathbf{q}}

\begin{document}

\title{Cavity-controlled chemistry in molecular ensembles}
\author{Felipe Herrera}
\affiliation{Department of Physics, Universidad de Santiago de Chile, Av. Ecuador 3943, Santiago, Chile}
\author{Frank C. Spano}
\affiliation{Department of Chemistry, Temple University, Philadelphia, Pennsylvania 19122, USA}

\date{\today}

\begin{abstract}  
The demonstration of strong and ultrastrong coupling regimes of cavity QED with polyatomic molecules has opened new routes to control chemical dynamics at the nanoscale. We show that strong resonant coupling of a cavity field with an electronic transition can effectively decouple collective electronic and nuclear degrees of freedom in a disordered molecular ensemble, even for molecules with high-frequency quantum vibrational modes having strong electron-vibration interactions. This type of polaron decoupling can be used to control chemical reactions. We show that the rate of electron transfer reactions in a cavity can be orders of magnitude larger than in free space, for a wide class of organic molecular species. 
\end{abstract}


\maketitle

The experimental realization of the strong \cite{Lidzey1998,Lidzey1999,Lidzey2000,Tischler2005,Holmes2007,kena-cohen2008,Kena-Cohen2010,Hutchison2013,Bellessa2014} and ultrastrong \cite{Schartz2011,Kena-Cohen2013,Mazzeo2014,Cacciola2014,George2015-farad,Gambino2015} coupling regimes of cavity quantum electrodynamics (QED) with organic matter in optical cavities has stimulated interest in the development of hybrid quantum devices with enhanced energy and electron transport properties \cite{Andrew2000,Feist2015,Schachenmayer2015,Orgiu2015}, tunable nonlinear optical response \cite{Herrera2014}, and novel optomechanics \cite{Hyun2013}. The strong resonant coupling between a cavity mode and electronic \cite{Tischler2007,Torma2015} or vibrational \cite{Long2015,Muallem2015,George2015} molecular transitions is well-known to result in polariton formation \cite{Mabuchi2002,Vahala2003,Litinskaya2006,Khitrova2006,RMiller2005,Torma2015}. 
However, the interplay between electronic and nuclear degrees of freedom under strong coupling with a cavity field remains widely unexplored \cite{Spano2015,Galego2015}. Chemistry is dominated by the coupling between electrons and vibrations. Therefore, it is important to understand the role that a cavity field can play to alter the electron-vibration dynamics, which would provide a path to control chemistry using cavity QED.


In this work, we show that the strong collective interaction of a molecular ensemble with the vacuum field of an optical cavity can in fact modify the nuclear dynamics of individual molecules in the ensemble. In free space, when an electron is optically excited, the nuclei in a molecule rearrange to a configuration that minimizes the electronic energy in the excited state. The excited nuclear configuration is in typically different from the ground state equilibrium configuration. We find that in an optical cavity that can exchange energy with a collective electronic state faster than the timescales associated with nuclear motions, reorganization of the nuclei upon excitation is strongly suppressed. This effect is a type of polaron decoupling involving collective electronic degrees of freedom that are symmetric with respect to molecular permutations. We show that polaron decoupling can occur in molecular ensembles with a large degree of energetic disorder, which is typical of organic systems. The effective manipulation of intramolecular nuclear dynamics can be used to control chemical reactivity, for example, by controlling the reorganization energy in Marcus electron transfer reactions \cite{Marcus1993}. In order to illustrate this possibility, we show that strong cavity-matter coupling can significantly enhance the rate of intramolecular electron transfer (ET) reactions within individual molecules in the ensemble. To the best of our knowledge, this is the first demonstration of collective photon-assisted control of local chemical properties.

\begin{figure}[t]
\centering
\includegraphics[width=0.45\textwidth]{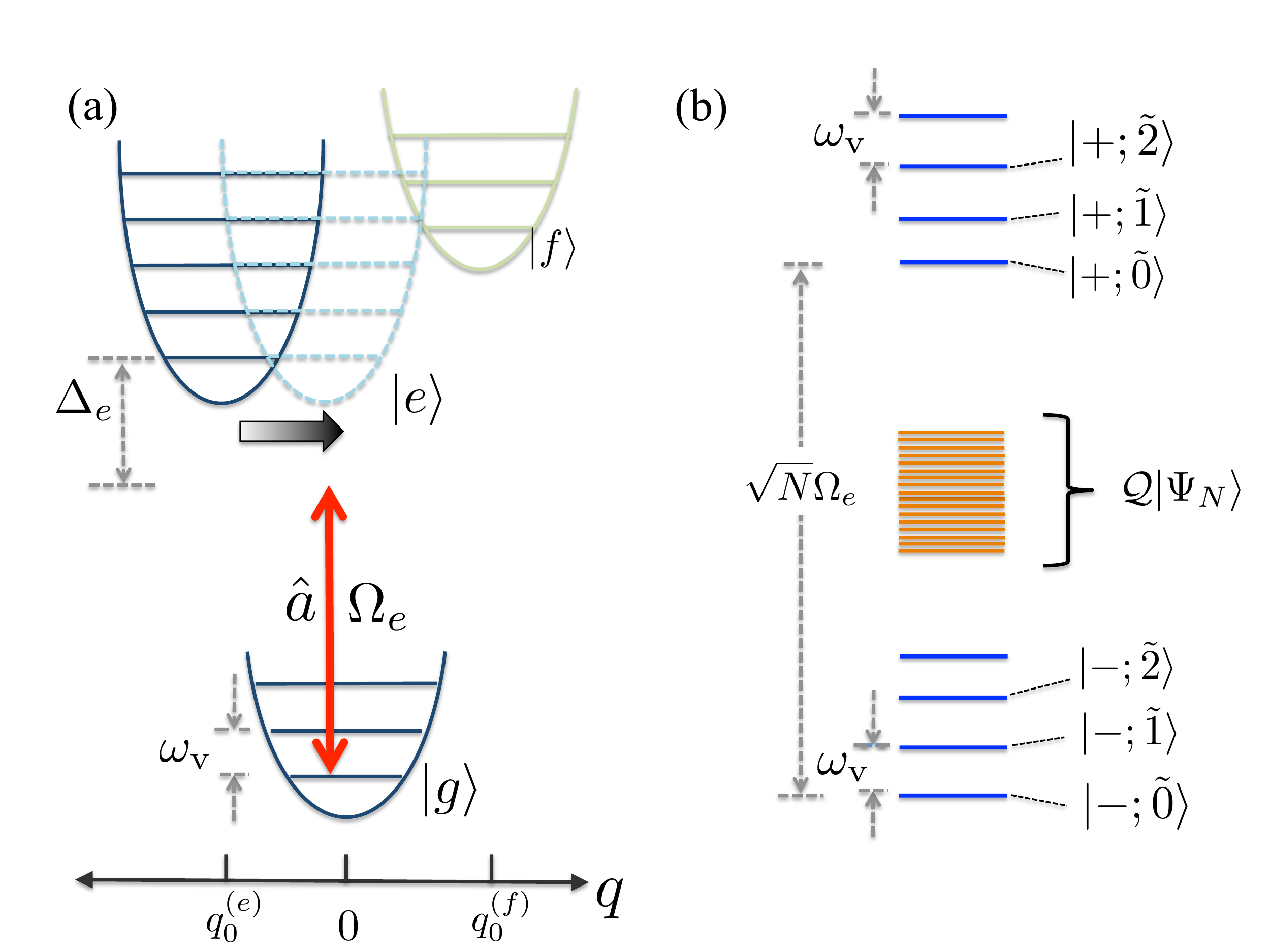}
\caption{Panel (a): Nuclear potentials for a single molecule along the reaction coordinate $q$. Excited states $\ket{e}$ and $\ket{f}$ are shifted with respect to the ground state by $q_0^{(e)}$ and $q_0^{(f)}$ in free space. A cavity field $\hat a$ couples to the transition $\ket{g}\leftrightarrow\ket{e}$ with a detuning $\Delta_e$ and Rabi frequency $\Omega_e$. For large enough $\Omega_{e}$, the cavity effectively shifts the potential minimum in state $\ket{e}$ to coincide with $\ket{g}$ (right grey arrow). State $\ket{f}$ does not couple to the cavity. Panel (b): Cavity-dressed spectrum for an ensemble of $N$ molecules with vibronic coupling, showing the splitting of the permutation-symmetric collective dressed states $\ket{\pm,\tilde m}$ from the cavity-free polaron eigenstates $\mathcal{Q}\ket{\Psi_N}$. $\omega_{\rm v}$ is the vibrational frequency.}
\label{fig:diagram}
\end{figure}

Polyatomic molecules with $z$ atoms have $3z - 6$ normal modes of vibration, each mode involving the coupled motion of multiple atoms within a molecule. Often only a few of these modes are needed to describe a chemical reaction \cite{Miller1980}. These are known as the reaction coordinates (RC), associated with a mass-weighted superposition of the form $q_k = \sum_i\alpha_{ik}\sqrt{m_i}x_i$, where $k$ is a mode index and $x_i$ is the displacement of the $i$-th atom from the potential minimum. Each atomic displacement has a mass-weighted momentum $p_i/\sqrt{m_i}$.
For a single reaction coordinate $q$ in the harmonic approximation for the potential, we can represent the ground state energy as $H_g(q) = (p^2 + \omega_{\rm v}^2q^2)/2$, where $\omega_{\rm v}$ is the frequency of the intramolecular vibration and $p$ is the normal mode momentum. 
The nuclei have a reference equilibrium configuration $q_0=0$ in the ground state $\ket{g}$. The equilibrium nuclear configuration in an excited electronic state is however different from the ground state configuration. This difference is a manifestation of electron-vibration coupling, or vibronic coupling. In the harmonic approximation, the nuclear potential in an excited state $\ket{e}$ is given by
\begin{equation}
H_e(q) = \omega_e+ \frac{1}{2}\left[p^2 +\omega_{\rm v}^2 (q - q_{0}^{(e)})^2\right],
\end{equation}
 where $\omega_e$ is the electronic energy and $q_{0}^{(e)}$ is a state-dependent shift that quantifies the degree of vibronic coupling. We set $\hbar = 1$ throughout and assume that the vibrational frequency is the same in all electronic states. In Fig. \ref{fig:diagram}a, we show the nuclear potentials for a molecule with a ground state $\ket{g}$ centered at $q_0=0$, as well as excited states $\ket{e}$ and $\ket{f}$, centered on opposite sides of the ground state minimum. This nuclear arrangement is relevant to describe electron transfer in substituted biphenyls \cite{Maus2002,Grabowski2003}, where the RC represents a torsional angle between the electron donor and acceptor groups. 
%
%
In Fig. \ref{fig:diagram}a, we also represent a quantized cavity mode $\hat a$ that induces Rabi oscillations between states $\ket{g}$ and $\ket{e}$ at frequency $\Omega_{\rm e}$ with detuning $|\Delta_e|\ll \Omega_e$.

Shifted harmonic oscillator modes $\tilde b$ for state $\ket{e}$ can be obtained from the unshifted ground state oscillator mode $b$ by a displacement along the RC, so that $\tilde b = b +\lambda_e$ \cite{May-Kuhn}. A similar definition holds for state $\ket{f}$. The dimensionless Huang-Rhys factor $\lambda_e^2 \equiv (\omega_{\rm v}/2)\;[q_{0}^{(e)}]^2$ characterizes the strength of vibronic coupling.
%
%
The vibrational eigenstates of the shifted harmonic potential are denoted by $\ket{\tilde m_e} \equiv D^\dagger(\lambda_e) \ket{m}$, where $D(\lambda_e) = {\rm exp}[\lambda_e(b^\dagger - b)]$ is the nuclear displacement operator and $\ket{m}$ is a vibrational eigenstate for the reference mode $b$. 

We are interested in the dynamics of an ensemble of $N$ identical molecules interacting with a single quantized electromagnetic mode of an optical cavity, having annihilation operator $\hat a$. 
 The many-body Hamiltonian for the ensemble can be written in the interaction picture as
\begin{eqnarray}\label{eq:HN}
\mathcal{H}_N&=&  \omega_{\rm v}\sum_\nu  b_\nu^\dagger  b_\nu +\sum_{\alpha\beta\nu}\left[\delta_{\alpha\beta}\Delta+g_{\alpha\beta}^\nu\left(b_\nu +b_\nu^\dagger\right)\right] \ket{\alpha}\bra{\beta} \nonumber\\
&&- i\sqrt{N}\left(\frac{\Omega_e}{2}\right)\left(\ket{\alpha_0}\bra{G}\,\hat a-\ket{G}\bra{\alpha_0} \,\hat a^\dagger\right),
\end{eqnarray}
where $\ket{G} = \ket{g_1g_2\ldots g_N}$ is the ensemble ground state, $\ket{\alpha} = \sum_i u_{\alpha i}\ket{e_i}$ are one-excitation states and $\hat b_\nu ^\dagger= \sum_{i} c_{\nu i} \,\hat b_i^\dagger$ creates a phonon in mode $\nu$. The detuning from the cavity frequency $\omega_{\rm cav}$ is $\Delta = \Delta_e +\omega_{\rm v}\lambda_e^2 $, with $\Delta_e = \omega_e - \omega_{\rm cav}$ is the detuning from the zero-phonon line (0-0) transition and $\Omega_e$ is the single-molecule Rabi frequency. We assume that $|\Delta_{e}|/\Omega_e \ll 1$. Other excited states (for example $\ket{f}$ in Fig. \ref{fig:diagram}a) are far detuned from the cavity field and do not exchange energy with the confined mode over the timescales of interest.

The electron-vibration coupling constant in eq. (\ref{eq:HN}) is given by $g_{\alpha\beta}^\nu = \lambda_e\omega_{\rm v} \sum_i u_{\alpha i}^* \,c_{\nu i}\,u_{\beta i}$. By defining the $\nu = 0$ phonon mode to be totally-symmetric with respect to particle permutations, i.e., $c_{0i} = N^{-1/2}$, we have $g_{\alpha\beta}^{\nu=0} = \delta_{\alpha,\beta}\lambda_e\omega_{\rm v}/\sqrt{N}$. In other words, the permutation-symmetric phonon mode does not couple different collective electronic states and therefore do not lead to polaron formation \cite{Holstein:1959}.

We refer to Hamiltonian $\mathcal{H}_N$ in eq. (\ref{eq:HN}) as a Holstein-Jaynes-Cummings (HJC) model. In free space we can set $\Omega_e\rightarrow 0$ and $\Delta_e\rightarrow\omega_e$ to recover a standard Holstein model with optical phonons \cite{Holstein:1959}, which is used to describe small polaron dynamics \cite{May-Kuhn,Holstein:1959,salje2005polarons}. It is straightforward to generalize eq. (\ref{eq:HN}) to include direct long-range interactions between molecules. For an ensemble with translational symmetry in a lattice, we identify the mode indices $\alpha$ and $\nu$ with the quasi-momenta $\k$ and $\q$ of electronic and phonon excitations, respectively. We use this $(\k,\q)$ lattice representation below for numerical diagonalization of eq. (\ref{eq:HN}).

The cavity field profile is assumed to be constant over the volume occupied by the molecular ensemble. Therefore, the cavity mode can exchange energy efficiently only with the permutation-symmetric electronic state $\ket{\alpha_0} = \sum_i \ket{e_i}/\sqrt{N}$ (second line eq. (\ref{eq:HN})), with a size-enhanced Rabi frequency $\sqrt{N}\Omega_e$. We show in the Supplementary Material (\href{link}{SM}) that we can exploit the selection of permutation-symmetric electronic states by the cavity to
%
partition the electronic Hilbert space into a symmetric subspace $\mathcal{P} = \ket{G}\bra{G}+\ket{\alpha_0}\bra{\alpha_0}$ and a non-symmetric manifold $\mathcal{Q} =1_N - \mathcal{P}$, where $1_N$ is the many-body identity. Equation (\ref{eq:HN}) can be projected into these orthogonal manifolds to give
%
%
\begin{equation}\label{eq:HN partitions}
\mathcal{H}_N = \mathcal{P}^\dagger \mathcal{H}_N \mathcal{P}+\mathcal{Q}^\dagger \mathcal{H}_N \mathcal{Q}+\mathcal{P}^\dagger \mathcal{H}_N \mathcal{Q}+{\rm H.c.}.
\end{equation}
The specific forms for each term in eq. (\ref{eq:HN partitions}) are given in the \href{link}{SM}.

Electron-vibration coupling in the $\mathcal{P}$ manifold involves only the symmetric phonon ($\nu = 0$)  with coupling constant $g_{\alpha_0\alpha_0}^{\nu=0}= \lambda_e\omega_{\rm v}/\sqrt{N}$. Non-symmetric phonon modes ($\nu\neq 0$) have the same equilibrium configuration in states $\ket{\alpha_0}$ and $\ket{G}$. Vibronic coupling in the symmetric mode is also suppressed by a factor of $1/\sqrt{N}$ with respect to single-molecule vibronic coupling, as discussed above. In the \href{link}{SM} we provide an equivalent derivation of the cavity-induced reduction of the Huang-Rhys factor for lattice vibrations using phonon modes in the site basis $b_i$. We choose the collective phonon basis $b_\nu$ here to make a clear connection with lattice systems that are relevant to describe molecular aggregates \cite{Herrera2014,Spano2015}. 

The permutation-symmetric electronic partition $\mathcal{P}^\dagger \mathcal{H}_N \mathcal{P}$ can  be diagonalized ($\Delta_e = 0$) to obtain dressed vibronic states of the form
\begin{equation}\label{eq:dressed eigenstates}
\ket{\pm; \tilde m} = \ket{\psi_\pm} \otimes \hat D^\dagger(\lambda_e/2\sqrt{N})\ket{m} \otimes \ket{\{m\}_{\nu'}},
\end{equation}
where $\ket{\psi_\pm} = \{\ket{G}\ket{n_{\rm cav} = 1}\pm \ket{\alpha_0} \ket{n_{\rm cav} = 0}\}/\sqrt{2}$ is the dressed state in the absence of vibrations. $\ket{n_{\rm cav}}$ is a Fock state of the cavity mode, which we restrict here to $n_{\rm cav}\leq 1$ to study cavity-vacuum effects only. State $\ket{m}$ is a vibrational eigenstate of the symmetric phonon mode $b_{\nu = 0}$ and $\ket{\{m\}_{\nu'}}\equiv \ket{m_{\nu =1},\ldots,m_{\nu=N-1}}$ describes the vibrational state in non-symmetric phonon modes. 
Dressed states in eq. (\ref{eq:dressed eigenstates}) have energies given by
\begin{equation}\label{eq:dressed energies}
\omega_{\pm,\tilde m} = \pm \sqrt{N}\Omega_e/2 + \omega_{\rm v}\kappa(\tilde m)+\omega_{\rm v}\sum_{\nu'\neq 0}m_{\nu'},
\end{equation}
where $\kappa(\tilde m)\approx \tilde m$ for low vibrational quanta. Stokes shifts of order $1/N$ are ignored. We illustrate the spectrum of the many-body Hamiltonian $\mathcal{H}_N$ in Fig. \ref{fig:diagram}b.
We find that as $\sqrt{N}\Omega_e/\omega_{\rm v}\gg 1$, coupling between the $\mathcal{P}$ and $\mathcal{Q}$ manifolds through phonon absorption or emission becomes strongly suppressed. In this regime, the many-body states given in eq. (\ref{eq:dressed eigenstates}) become eigenstates of the HJC model with energies given by eq. (\ref{eq:dressed energies}). 
We show below that this form of polaron decoupling can have a significant impact on the chemical reactivity of molecular ensembles in optical cavities.

\begin{figure}[t]
\centering
\includegraphics[width=0.48\textwidth]{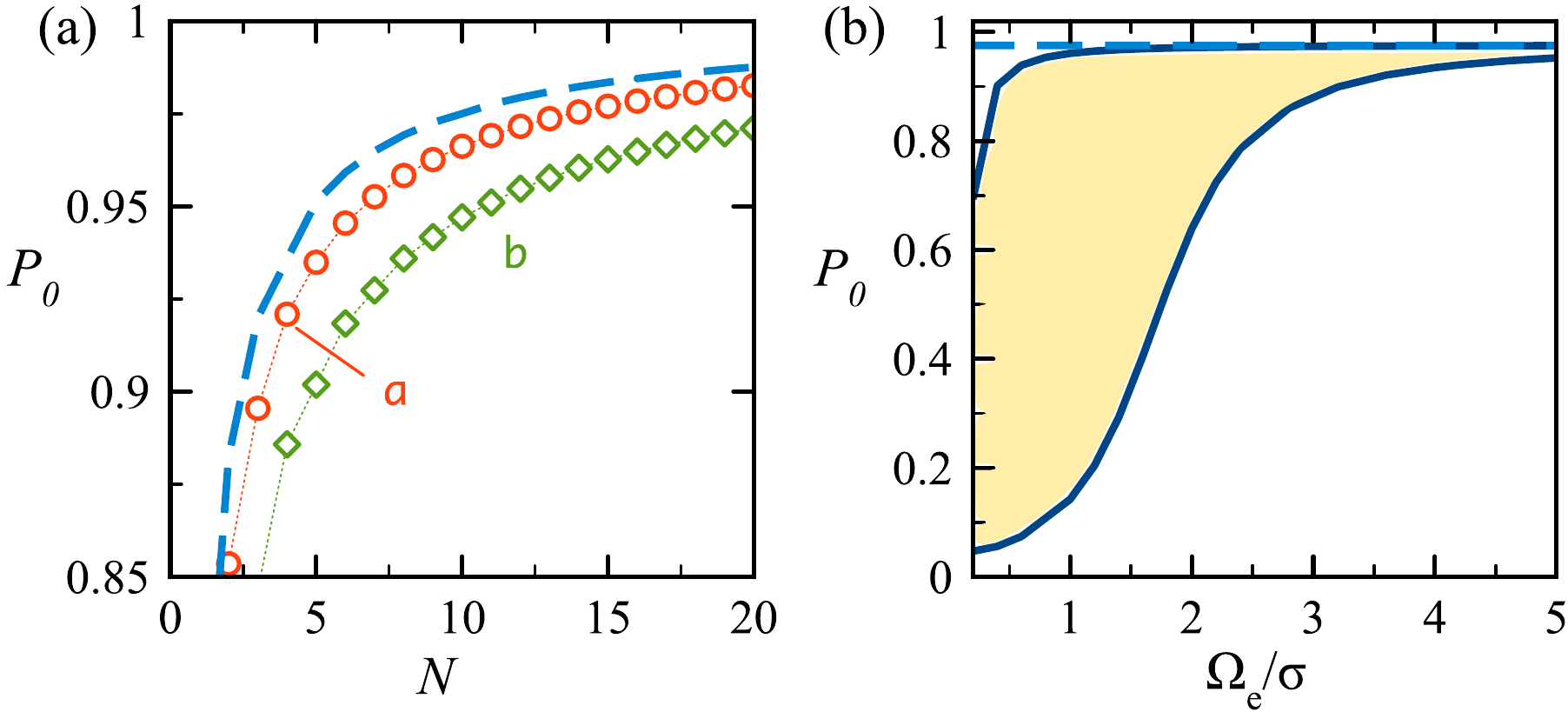}
\caption{Cavity-induced polaron decoupling. {Panel (a)}: Probability $P_0$ for the lowest many-body dressed state to be decoupled from molecular vibrations, as a function of the number of molecules $N$ in a linear lattice. Several values of the Rabi frequency are shown: $\Omega_{\rm e} =4$ (curve $a$) and $\Omega_{\rm e} = 2$ (curve $b$).
Panel (b): $P_0$ as a function of $\Omega_{e}/\sigma$ for a linear array of size $N = 10$ and disorder width $\sigma$. 2500 disorder realizations are included in the shaded area.  Dashed lines in both panels correspond to $P_0 = {\rm exp}[-\lambda_{\rm e}^2/4N]$, with $\lambda_e^2 = 1$.} 
\label{fig:state projection}
\end{figure}

We illustrate the phenomenon of polaron decoupling in Fig. \ref{fig:state projection}. We quantify the degree of vibronic coupling in the eigenstates of the HJC model by the squared-overlap $P_0 \equiv |\langle\Phi_0 \ket{\psi_-;m=0}\rangle |^2$, where $\ket{\Phi_0}$ is the lowest polariton eigenstate of $\mathcal{H}_N$ for a resonant cavity ($\Delta_e = 0$), and $\ket{\psi_-}\otimes \ket{m=0}$ is given by eq. (\ref{eq:dressed eigenstates}) for $\lambda_e = 0$. As we mentioned above, periodic boundary conditions are used to represent collective electronic and vibrational states, including up to six vibrational quanta in the symmetric mode $b_0$ and up to two in the non-symmetric modes $ b_{\nu\neq 0}$.  
We find that the squared-overlap satisfies $P_0\leq {\rm exp}[-\lambda_e^2/4N]$ (see \href{link}{SM} for a derivation). The upper bound corresponds an effective Franck-Condon overlap with Huang-Rhys factor $\lambda_e^2/4N$ between the collective nuclear states in state $\ket{\alpha_0}$ and those in state $\ket{G}$. This overlap becomes exponentially close to unity in the limit $N\gg \lambda_e^2$. The nuclear configuration of the collective excited electronic state $\ket{\alpha_0}$ is thus the same as the ground state equilibrium configuration, for all phonon modes $\nu$. In other words, when the collective Rabi oscillation period is shorter than the timescales for vibrational motion, the electron can exchange energy with the cavity mode many times before the nuclei have time to reorganize their configuration to the excited state potential.

We now go beyond the restriction of identical molecules and consider an excited state energy $\omega_{e}(\r_i)$ that depends on the position of the molecule $\r_i$ in the ensemble. This type of static disorder in organic systems is typically taken into account by assuming that $\omega_e(\r_i)$ has a Gaussian distribution with standard deviation $\sigma$ \cite{Agranovich2003,Spano2015}. For a linear array with $N=10$ molecules, we show in Fig. \ref{fig:state projection}b  that the lowest eigenstate $\ket{\Phi_0}$ of the HJC Hamiltonian $\mathcal{H}_N$ with random detunings $\Delta_e(\r_i)$ is accurately given by $\ket{-;\tilde 0}$ as $\sqrt{N}\Omega_e\gg {\sigma}$. In this limit, the upper bound for $P_0$ becomes tight for all disorder realizations, as observed by the narrowing of the distribution in Fig. \ref{fig:state projection}b for $\Omega_e/\sigma\gg 1$. 

Having described polaron decoupling by energetic isolation of the permutation-symmetric $\mathcal{P}$-manifold, we now consider its effect on non-adiabatic unimolecular electron transfer (ET) reactions. In ET reactions, an excess electron is transferred from a donor (D) to an acceptor (A) group within a molecule. The coherent transfer rate $V$ is proportional to the orbital overlap between the D and A groups \cite{May-Kuhn,Marcus1993}. For non-adiabatic ET reactions we have $V/\hbar\ll \omega_{\rm v}$. 

The relative energy between donor and acceptor vibronic levels $\Delta E \equiv \omega_{DA} + (m_D-m_A)\omega_{\rm v}$ is known as the driving force of an ET reaction \cite{May-Kuhn}. $\omega_{DA} = \omega_D-\omega_A$ is the electronic transition frequency and $(m_D-m_A)\omega_{\rm v}$ the vibrational transition frequency for the D-A pair. The reaction rate can be written using linear response theory as \cite{May-Kuhn} ($\hbar = 1$)
\begin{equation}\label{eq:ET rate}
k_{\rm ET}(\Delta E) = 2\pi V^2 \sum_{m_D}\sum_{m_A}\eta_{m_D}(T)\mathcal{D}(\Delta E),
\end{equation}
where $\eta_{m_D}(T)$ is the Boltzmann distribution of nuclear states in the donor manifold $\ket{m_D}$ at temperature $T$ and $\mathcal{D}(\Delta E)$ is a Franck-Condon weighted lineshape function, whose specific form depends on the model assumed for the system-environment interaction and the relative donor-acceptor shift $\lambda_{DA} \equiv \lambda_D-\lambda_A$. For a closed system, we have $\mathcal{D}(\Delta E) = |\langle m_D|m_A\rangle|^2\delta(\Delta E)$, which gives Fermi Golden Rule. We are interested in high-frequency internal vibrations for which $k_{\rm B}T/\hbar\omega_{\rm v}\ll 1$ at room temperature. In this case, ET reactions occurs through nuclear tunneling and the rate is strongly suppressed away from the resonance condition $\Delta E = 0$ \cite{May-Kuhn}.

We consider a cavity-driven ET reaction in an ensemble of $N$ donor-acceptor complexes. The state $\ket{e}$ from Fig. \ref{fig:diagram} takes the role of the donor and state $\ket{f}$ becomes the acceptor. Outside the cavity, the ET reaction channel $\ket{e}\rightarrow \ket{f}$ is strongly suppressed when $\Delta E$ is away from a vibrational resonance between the donor and acceptor potentials. Inside the cavity, a confined photon can exchange energy resonantly with the transition $\ket{g}\leftrightarrow\ket{e}$ at frequency $\Omega_{\rm e}$, which delocalizes the donor state over the ensemble. On the other hand, the acceptor state $\ket{f}$ is far detuned to the blue of the cavity frequency and remains localized in each molecule. For $\sqrt{N}\Omega_{\rm e}\gg \omega_{\rm v}$, the donor states undergo polaron decoupling and can be represented by eq. (\ref{eq:dressed eigenstates}). 

We can open the ET reaction channel $\ket{e}\rightarrow \ket{f}$, by Rabi splitting the energy of the upper dressed donor state $\ket{+; \tilde m}$ above the acceptor level. $\omega_D=\sqrt{N}\Omega_e/2$ is the  donor electronic energy in a frame rotating at the cavity frequency $\omega$. We can thus have $|\Delta E|\ll  2\gamma_{\rm v}$ for a given donor vibrational level $\tilde m$, where $\gamma_{\rm v}$ is the vibrational relaxation rate. An electron placed in the dressed donor state $\ket{+;\tilde m}$ by an electron beam or a weak laser probe, is thus transferred to the acceptor state $\ket{f}$ at the rate $k_{\rm ET}(\Delta E)$. In an optical cavity, the rate expression in eq. (\ref{eq:ET rate}) is still valid, but the lineshape function is not the same as in free space. In the \href{link}{SM}, we derive the lineshape function for ET reactions in a cavity.

\begin{figure}[t]
\centering
\includegraphics[width=0.48\textwidth]{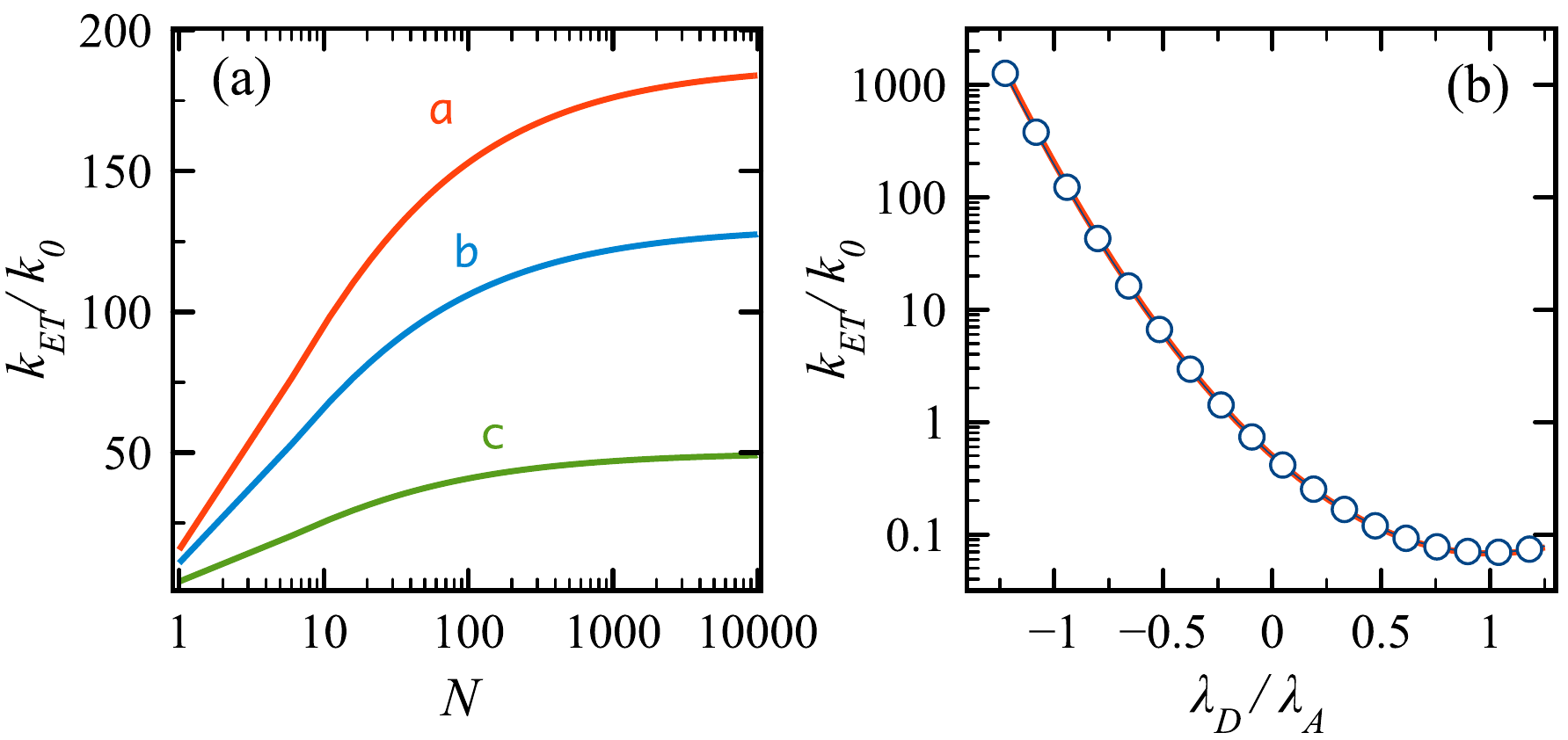}
\caption{Electron transfer (ET) rate $k_{\rm ET}$ in an optical cavity. Panel (a): Ratio $k_{\rm ET}/k_0$ as a function of $N$. $k_0$ is the ET rate in free space. Curves are shown for $\Delta E = 0$ (curve $a$), $\Delta E = 2\gamma_{\rm v}$ (curve $b$) and $\Delta E = 5\gamma_{\rm v}$ (curve $c$). We set $k_{\rm B}T/\hbar = 0.1$ and $\lambda_D = -\lambda_A = \sqrt{2}$. 
Panel (b): $k_{\rm ET}/k_0$ as a function of $\lambda_D/\lambda_A$ for $N = 10^4$ molecules with $\lambda_A = \sqrt{2}$ and $\Delta E = 0$ (circles). The solid line is the analytical bound in eq. (\ref{eq:ratio bound}). In both panels the vibrational relaxation rate is $\gamma_{\rm v} = 0.01 \,\omega_{\rm v}$, where $\omega_{\rm v}$ the vibrational frequency, and $k_{\rm T}T = 0.1 \omega_{\rm v}$. $\Delta E\ll \omega_{\rm v}$ is the donor-acceptor electronic transition frequency, taken to be the same in the cavity and in free space.}
\label{fig:ET rate}
\end{figure}

In general, the cavity-donor coupling has two main effects on the ET reaction rate $k_{\rm ET}$. The first effect is the Rabi splitting of the dressed donor energy $\omega_D$ relative to the acceptor levels. This energy shift can change the driving force of the reaction $\Delta E$ relative to its free-space value $\Delta E_0$. The cavity can thus resonantly enhance nuclear tunneling for ET reactions involving high-frequency modes. 

The second effect of cavity-donor coupling on ET reactions is related to polaron decoupling of the donor electronic state. As we discussed above, the collective coupling of donor groups to the same cavity mode effectively preserves the nuclear configuration of the ground electronic state along the collective reaction coordinate. For donor and acceptor excited states that in free space have equilibrium nuclear configurations with shifts of opposite signs ($\lambda_D \lambda_A<0$) relative to the $\ket{g}$, as in Fig. \ref{fig:diagram}a, polaron decoupling at low temperatures (or high vibrational frequencies) can increase the ET rate by orders of magnitude compared to the free space rate $k_0$. This is illustrated in Fig. \ref{fig:ET rate}a, where we show the ratio $k_{\rm ET}/k_0$ as a function of $N$ for fixed $\Delta E$. We assume that $\Delta E$ is the same in the cavity and in free space, which can be achieved by tuning the acceptor energy $\omega_A$ through chemical substitution \cite{Closs:1988} or by changing the solvent polarity \cite{Maroncelli:1989}. For low vibrational temperature, only $m_D = 0$ and $m_A = 0$ contribute to the rate for $\Delta E \ll \omega_{\rm v}$. In this case, the cavity rate $k_{\rm ET}$ for resonant tunneling ($\Delta E  =0$) can be approximated by
\begin{equation}\label{eq:ratio bound}
k_{\rm ET}= \left({k_0}/{2}\right)\,{\rm exp}[{\lambda_{D}^2-2\lambda_D\lambda_A}],
\end{equation} 
where $k_0$ is the resonant tunneling rate in free space. This expression is valid for $N\gg \lambda_D^2$. The ratio $k_{\rm ET}/k_0$ therefore exceeds unity for $\lambda_A\lambda_D<0$, but can be smaller than unity for donor and acceptor levels shifted in the same direction relative to $\ket{g}$. This is illustrated in Fig. \ref{fig:ET rate}b.


In summary, we discuss a mechanism for cavity-assisted decoupling of the nuclear and the electronic molecular degrees of freedom in a molecular ensemble. This type of polaron decoupling can enhance or suppress the rate of intramolecular electron transfer by orders of magnitude in comparison with free space. Since we only assume conditions of strong coupling of a single cavity mode with an electronic transition, our results are valid for organic systems in microcavities \cite{Holmes2007} and plasmonic nanocavities \cite{Gonzalez2013,Bellessa2014}. The predicted enhancements should be observable for a wide class of electron transfer reactions that involve large nuclear rearrangements in excited electronic states \cite{Grabowski2003}. In addition to intramolecular electron transfer, cavity-induced polaron decoupling can also be used to control bimolecular electron or energy transfer processes that involve nuclear rearrangements in excited electronic states, including F\"{o}rster resonance energy transfer \cite{Andrew2000}.

\acknowledgments
FH is supported by CONICYT through PAI 79140030 and Fondecyt Iniciaci\'{o}n 11140158. FCS is supported by the NSF,  grant \# DMR-1505437.

\bibliography{cavitychem}

\end{document}